\documentclass[aps,prl,twocolumn]{revtex4-1}

\usepackage{amsmath}
\usepackage{amssymb}
\usepackage{mathtools}
\usepackage{graphicx}
\usepackage[usenames,dvipsnames]{color}
\usepackage{bm}
\usepackage{hyperref}
\usepackage{url}
\usepackage[utf8]{inputenc}
\usepackage{subfigure}
\usepackage{slashed,bbm}
\usepackage{graphics,psfrag,epsfig}
\usepackage{dsfont}
\usepackage{setspace}
\usepackage{soul}
\usepackage{times}

\usepackage{geometry}
\makeatletter
\newcommand*{\toccontents}{\@starttoc{toc}}
\makeatother

\usepackage{hyperref}
\usepackage{xcolor}
\usepackage{times}
\definecolor{dark-red}{rgb}{0.4,0.15,0.15}
\definecolor{dark-blue}{rgb}{0.15,0.15,0.4}
\definecolor{medium-blue}{rgb}{0,0,0.5}
\hypersetup{
colorlinks, linkcolor={dark-blue},
citecolor={dark-blue}, urlcolor={medium-blue}
}

\begin{document}

\title{Emergence and stability of spin-valley entangled quantum liquids in moir\'e heterostructures}

\author{Dominik Kiese, Finn Lasse Buessen, Ciar\'an Hickey, Simon Trebst, and Michael M. Scherer}
\affiliation{Institute for Theoretical Physics, University of Cologne, 50937 Cologne, Germany}
\date{\today}

\begin{abstract}
	Twisting moir\'e heterostructures to the flatband regime allows for  the formation of strongly correlated quantum states, since the dramatic
	reduction of the bandwidth can cause the residual electronic interactions to set the principal energy scale.
	An effective description for such correlated moir\'e heterostructures, derived in the strong-coupling limit  at integer filling, generically leads to 
	spin-valley Heisenberg models. 	
	Here we explore the emergence and stability of spin liquid behavior in an 
	{SU(2)}$^{\mathrm{spin}}\otimes$ {SU(2)}$^{\mathrm{valley}}$ Heisenberg model 
	upon inclusion of Hund's-induced and longer-ranged exchange couplings, 
	employing a pseudofermion functional renormalization group approach. 
	We consider two lattice geometries, triangular and honeycomb (relevant to different moir\'e heterostructures), 
	and find, for both cases, an extended parameter regime surrounding the SU(4) symmetric point where no long-range  order occurs, 
	indicating a stable realm of quantum spin liquid behavior.
	For large Hund's coupling, we identify the adjacent magnetic orders,
	with both  antiferromagnetic and ferromagnetic ground states emerging in the separate spin and valley degrees of freedom.
	For both lattice geometries the inclusion of longer-ranged exchange couplings 
	is found to have both stabilizing and destabilizing effects on the spin liquid regime depending on the sign of the additional couplings.	
\end{abstract}

\maketitle

Spurred by the discovery of a plethora of  insulating and superconducting states in twisted bilayer graphene (TBG) \cite{cao2018correlated,cao2018unconventional}, a growing stream of experimental evidence points to the generic emergence of correlated electronic behavior in various moir\'e heterostructures~\cite{yankowitz2019tuning,sharpe2019emergent,lu2019superconductors,chen2019evidence,chen2019signatures,codecido2019correlated,liu2019spin,cao2019electric,shen2019observation,2019arXiv190700261S}.
The basic mechanism that gives rise to strongly-enhanced correlation effects in these materials is the formation of long-wavelength moir\'e patterns
with (almost) flat low-energy bands
whose narrow bandwidth becomes comparable to the otherwise negligible energy scale of the electronic interactions~\cite{doi:10.1021/nl902948m,Bistritzer12233,PhysRevB.86.155449}.
Due to a high degree of control, e.g., in the regulation of the twist angle, tunable bandwidths or fillings, and a low level of disorder, such systems are discussed as ideal platforms for  detailed studies of quantum many-body states.
Despite a vast amount of concomitant theoretical activity~\cite{PhysRevLett.121.087001,PhysRevB.98.045103,PhysRevB.98.075154,PhysRevB.98.241407,tang2018spin,PhysRevX.8.041041,PhysRevB.98.214521,lin2019chiral,PhysRevB.98.235158,huang2018antiferromagnetically,PhysRevB.98.085436,PhysRevB.97.235453,roy2018unconventional,PhysRevLett.121.217001,da2019magic,2018arXiv180506867Y,PhysRevB.98.205151,PhysRevLett.122.246401,PhysRevB.99.195120,seo2019ferromagnetic}, the precise nature of the observed insulators and superconductors, however, remains to be explored and settled through the construction of faithful models and application of appropriate quantum {\em many-body} approaches.

Several model constructions for correlated moir\'e materials have been put forward in terms of effective tight-binding descriptions on the moir\'e superlattice, augmented by various interaction terms~\cite{PhysRevX.8.031087,PhysRevX.8.031088,PhysRevX.8.031089}. 
Whereas details of the models may differ, they feature a series of universal traits:
(1)~an emergent hexagonal superlattice, 
(2)~a multi-orbital structure, and 
(3)~Hund's and extended Hubbard interactions.
More specifically, while TBG is preferably described using a honeycomb superlattice~\cite{PhysRevX.8.031087,PhysRevX.8.031088,PhysRevX.8.031089}, related structures such as  twisted double-bilayer graphene (TDBG) or trilayer graphene/hexagonal boron nitride heterostructures (TLG/h-BN)  are better captured by a triangular superlattice~\cite{chen2019evidence,PhysRevX.8.031089,2018arXiv181208097X}. The orbital degrees of freedom are inherited from the valleys in the original bands, e.g., the two Dirac valleys in the Brillouin zone of graphene.

These universal aspects can be combined into a minimal model, with a two-orbital extended Hubbard model~\cite{PhysRevLett.121.087001,PhysRevB.98.045103} serving as a paradigmatic starting point.
Its kinetic term $H_{t}\!=\!-t\sum_{\langle ij\rangle}\sum_{\alpha=1}^4(c_{i\alpha}^\dagger c_{j\alpha}^{\phantom\dagger}+\mathrm{h.c.})$ 
for the electrons combines the spin projection $s \in \{ \uparrow,\downarrow\}$ and valley quantum number $l \in \{+, -\}$  in a flavor index $\alpha\in\{(\uparrow,\!+),(\uparrow,\!-),(\downarrow,\!+),(\downarrow,\!-)\}$, reflecting an effective SU(4) symmetry. On the triangular lattice this results in a set of four degenerate bands, which can describe, e.g., the set of minibands above charge neutrality in TDBG or TLG/h-BN. On the honeycomb lattice, with its additional sublattice degree of freedom, this results in eight bands, two sets of four degenerate bands, which describe the minibands above and below charge neutrality in TBG~\cite{SM}.

The simplest conceivable interaction term, which also retains the SU(4) symmetry, is a Hubbard interaction $H_{\mathrm{int}}=U\sum_i(\sum_{\alpha=1}^4 n_{i\alpha})^2$, 
which can arise in the limit of large lattice periods where the interaction depends primarily on the total charge on a site and becomes the dominant interaction scale.
In this strong-coupling limit, the kinetic term can then be treated perturbatively~\cite{PhysRevLett.121.087001,PhysRevB.98.045103,PhysRevX.8.031089}.  With an integer number of electrons per site this leads to an effective  spin-valley  Heisenberg Hamiltonian  with SU(4) symmetric superexchange coupling $J_H\propto t^2/U$. 
Additional symmetry-breaking interactions are, however, to be expected, in particular in the form of Hund's-type couplings  in either the spin or  valley degrees of freedom~\cite{PhysRevLett.121.087001,PhysRevB.98.045103}. 
Moreover, Wannier state constructions suggest that further-neighbor interactions can become sizable~\cite{PhysRevX.8.031087} and should augment any minimal model. 

In this work, the above considerations naturally lead us to explore a nearest-neighbor spin-valley Heisenberg model with $\mathrm{SU(2)}^{\mathrm{spin}}\otimes \mathrm{SU(2)}^{\mathrm{valley}}$ symmetry for both triangular and honeycomb lattice geometries, which we later supplement with further-neighbor interaction terms. Our focus is on the case of half-filling of the underlying Hubbard model, i.e.~two electrons per site. For the effective Heisenberg model at strong coupling, this implies that we are working with the six-dimensional self-conjugate representation of SU(4) spins. This is in contrast to the four-dimensional fundamental representation of SU(4) relevant to, e.g., the case of quarter-filling.

For both lattice geometries, we find extended parameter regimes surrounding the SU(4) symmetric point where no long-range symmetry-breaking order occurs, indicating a stable realm for a spin-valley entangled quantum liquid. 
Moving further away from the SU(4) symmetric point, we find magnetic order in the spin and valley degrees of freedom that can be either antiferromagnetic or ferromagnetic, providing a possible explanation for the clear signatures of spin-polarization observed in TDBG~\cite{liu2019spin,cao2019electric,shen2019observation}. 
To explore the effect of longer-range interactions, we augment our model by a next-nearest neighbor coupling and determine its role in stabilizing quantum spin-valley liquid (QSVL) behavior versus long-range order for different signs of the coupling and the two lattice geometries.
Our work complements earlier work for the case of quarter-filling, 
where it was argued that a QSVL state with neutral gapless fermionic excitations forms on the honeycomb lattice~\cite{PhysRevX.2.041013}, while on the triangular lattice extended parameter regimes without any net magnetization have been identified in DMRG simulations~\cite{wu2019ferromagnetism}.

\paragraph{Spin-valley model.} 

The starting point of our study is an SU(4) spin-valley Heisenberg model~\cite{PhysRevLett.121.087001,PhysRevX.8.031089},
$\mathcal{H}_{\mathrm{SU(4)}} = J_H \sum_{\langle ij \rangle} \hat{T}^\mu_{i} \hat{T}^\mu_{j}$, 
where $J_H $ is the antiferromagnetic exchange coupling between nearest neighbors on either the triangular or honeycomb lattice, and $\hat{T}_{i}$ denote SU(4) spins.
The $\mu\!=\!1,\dots,15$ components of the spin operators can be represented on a fermionic Hilbert space via the parton construction $\hat T^{\mu}_{i} = f^{\dagger}_{i \alpha} T^{\mu}_{\alpha \beta} f^{\phantom{\dagger}}_{i \beta}$, where the index $\alpha$ enumerates four different fermion flavors and the matrices $T^{\mu}$ are the SU(4) generators~\cite{Arovas1988}. 
At half-filling of the underlying Hubbard model, the local spin-valley Hilbert space is six-dimensional (4 choose 2), which leads to a local filling constraint of two partons per lattice site $\sum_\alpha f^\dagger_{i\alpha}f^{\phantom{\dagger}}_{i\alpha}\!=2$.
\begin{figure}[t]
  \centering
  \includegraphics[width = 1.0\columnwidth]{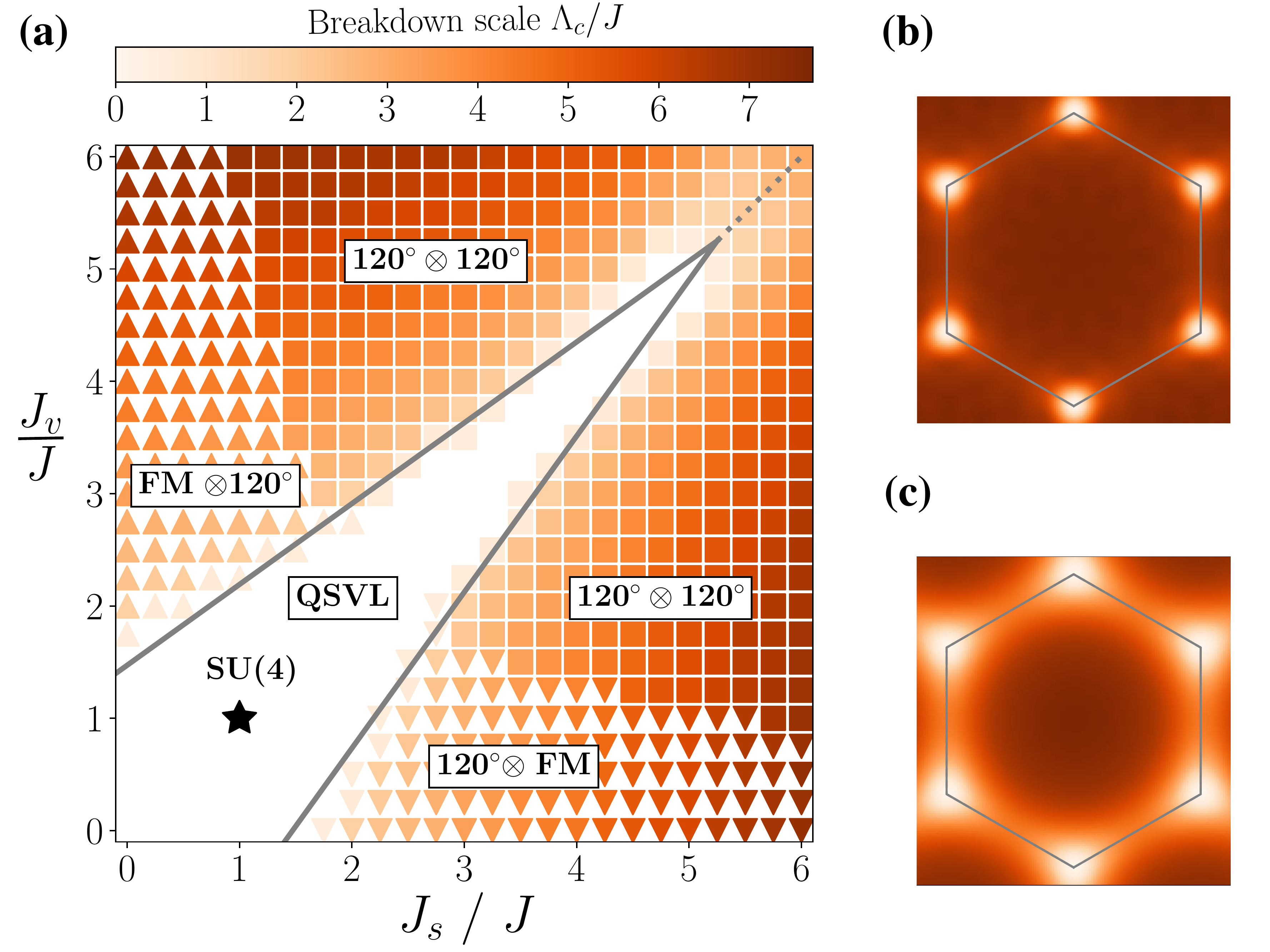}
\caption{{\bf Phase diagram on the triangular lattice.} 
  		(a) Colors indicate the magnitude of the breakdown scale $\Lambda_{c}$ in units of~$J$, triangles (squares) denote regions with negative (positive) effective coupling \eqref{eq:jeff}, see text for details. 
  		(b) Structure factor in the spin (valley) subspace at dominant $J_s$ ($J_v$), plotted at $\Lambda_{c}$, indicating the onset of $120^{\circ}$ order. 
  		(c) Structure factor at the SU(4) point where no instability of the RG flow occurs. Local correlations are reminiscent of $120^{\circ}$ order albeit broadened. The same color scale is applied to both (b) and~(c).
  	The solid gray lines mark the phase boundaries between the QSVL and the ordered phases, the dotted line marks the diagonal $J_s/J=J_v/J$.}
  \label{Fig:PhaseDiagramTriangular}
\end{figure}
Upon inclusion of Hund's couplings, the SU(4) symmetry of the model is explicitly broken~\cite{PhysRevLett.121.087001}. 
Omitting other sources of SU(4) breaking, a residual separate spin-valley $\mathrm{SU(2)}^{\rm s}\otimes \mathrm{SU(2)}^{\rm v}$ symmetry remains which is reflected by the extended Hamiltonian 
\begin{align}\label{eq:H}
\mathcal{H} &\!=\! \sum_{\langle ij \rangle} J(\hat{\sigma}^{a}_{i}\! \otimes\! \hat{\tau}^{b}_{i})(\hat{\sigma}^{a}_{j}\! \otimes\! \hat{\tau}^{b}_{j})\!+\! J_{s} \hat{\sigma}^{a}_{i} \hat{\sigma}^{a}_{j}
\!+\! J_{v} \hat{\tau}^{b}_{i} \hat{\tau}^{b}_{j}\,,
\end{align}
where the spin-valley operators read
$\hat{\sigma}^{a}_{i} = f^{\dagger}_{is'l'} \theta^{a}_{s's} \delta_{l'l} f^{\phantom{\dagger}}_{isl}$,
$\hat{\tau}^{b}_{i} = f^{\dagger}_{is'l'} \delta_{s's} \theta^{b}_{l'l} f^{\phantom{\dagger}}_{isl}$, and
$\hat{\sigma}^{a}_{i} \otimes \hat{\tau}^{b}_{i} =$ $f^{\dagger}_{is'l'} \theta^{a}_{s's} \theta^{b}_{l'l} f^{\phantom{\dagger}}_{isl}$.
Instead of enumerating the four fermion types by a single index, we have exposed the spin quantum number $s \in \{ \uparrow, \downarrow\}$ and the valley quantum number $l \in \{+, -\}$ explicitly; Pauli matrices are denoted by $\theta^{a}$, $a \in \{1,2,3\}$. 
At the high-symmetry point $J=J_{s}=J_{v}$ the full SU(4) symmetry is restored. We assume that the Hund's interactions are weak enough such that all exchange couplings are antiferromagnetic~\cite{wu2019ferromagnetism}, i.e. $J, J_v, J_s>0$.

\paragraph{Pseudofermion functional RG.}

Parton-decomposed quartic Hamiltonians of the general type defined in Eq.~\eqref{eq:H} can readily be analyzed by 
the pseudofermion functional renormalization group (pf-FRG)~\cite{PhysRevB.81.144410,WETTERICH199390}. For SU($N$) spins, the approach is already naturally formulated with a 
local constraint of $N/2$ fermions per site. 
It combines aspects of an expansion in spin length~$S$~\cite{PhysRevB.96.045144} (which naturally favors magnetic order) and in the SU($N$) spin symmetry~\cite{PhysRevB.97.064415,PhysRevB.97.064416} (which typically favors quantum spin liquid states), and it becomes exact on a mean-field level in the separate limits of large $S$ and large $N$.
It is thus suited to resolve the competition between ordered ground states and QSVL phases in the spin-valley model at hand. 
We extend the standard implementation of pf-FRG to incorporate the $\mathrm{SU(2)}^{\rm s}\otimes \mathrm{SU(2)}^{\rm v}$ symmetry, thereby obtaining flow equations for the one-particle irreducible vertices as a function of an RG frequency cutoff scale $\Lambda$~\cite{SM}. Numerically solving the set of $\mathcal{O}(10^6)$ flow equations at up to $84$ Matsubara frequencies and using a real-space vertex truncation of $L\!=\!7$ lattice bonds in each spatial direction, spontaneous symmetry breaking, e.g., the onset of long-range magnetic or valence bond order, is indicated by an instability of the RG flow~\cite{PhysRevB.81.144410,roscher2019cluster}  
which occurs at some critical scale~$\Lambda_c$.

\begin{figure}[t]
  \centering
  \includegraphics[width = 1.0\columnwidth]{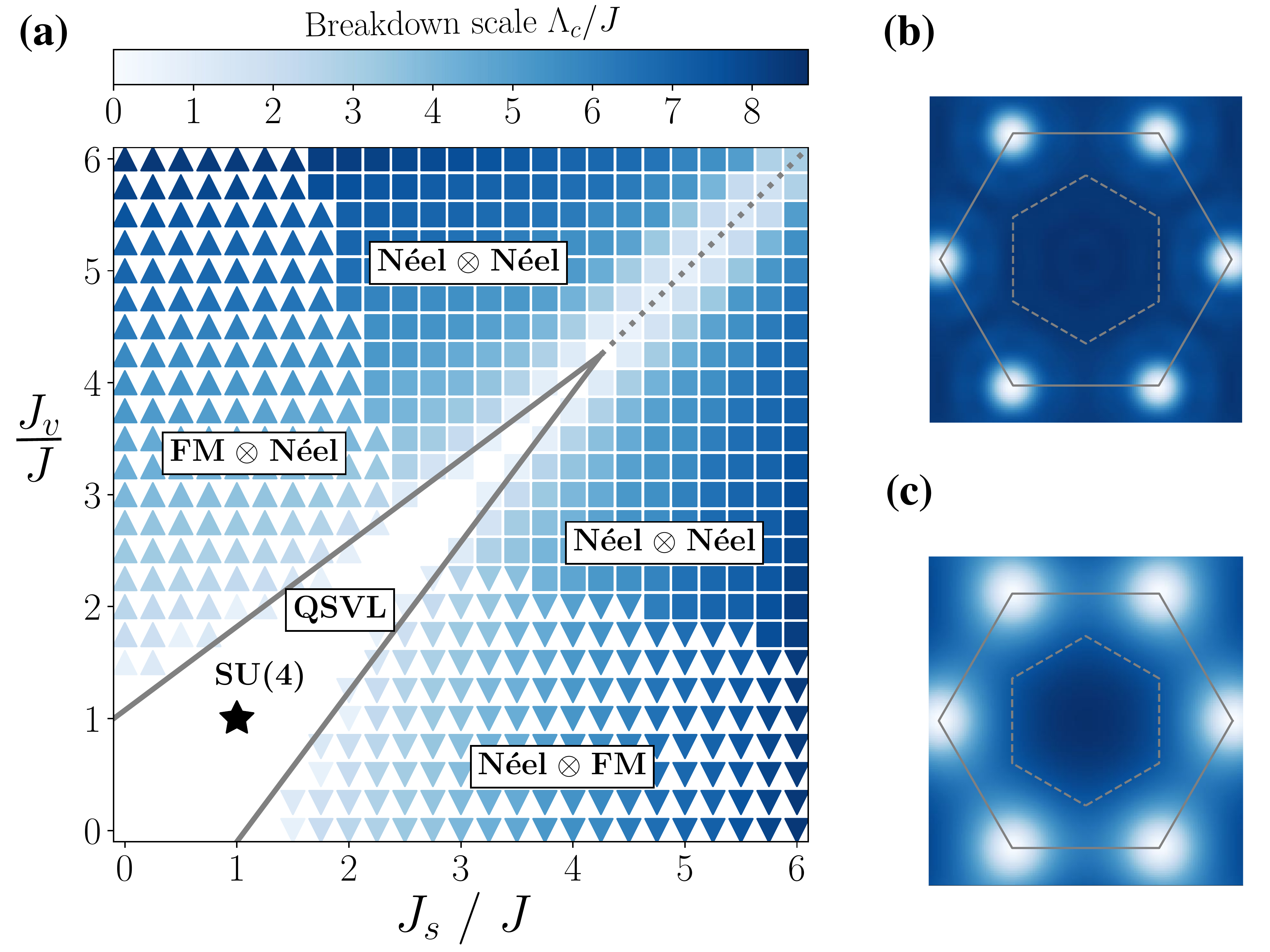}
  \caption{{\bf Phase diagram on the honeycomb lattice.}
  		(a) Colors indicate the magnitude of the breakdown scale $\Lambda_{c}$ in units of~$J$, triangles (squares) denote regions with negative (positive) effective coupling \eqref{eq:jeff}, see text for details.
		(b), (c) Structure factors for a state deep in the N\'eel ordered phase versus the SU(4) symmetric state,
		 with the same color scale applied.
			}
  \label{Fig:PhaseDiagramHoneycomb}
\end{figure}

In the case of long-range order, to identify the precise nature of the ordered state we can separately gain access to the elastic component ($\omega=0$) of the correlation functions in the spin sector and in the valley sector, 
\begin{align}
\chi_{ij}^{s \Lambda} = \langle \hat{\sigma}^{a}_{i} \hat{\sigma}^{a}_{j} \rangle^{\Lambda}\,,\quad\mathrm{and/or}\quad
\chi_{ij}^{v \Lambda} = \langle \hat{\tau}^{b}_{i} \hat{\tau}^{b}_{j} \rangle^{\Lambda}\,.
\end{align}
Sharp features emerging in the respective structure factors $\chi^{s/v}(\vec{q})\propto\sum_{ij}e^{i\vec{q}\cdot(\vec{r_i}-\vec{r_j})}\chi_{ij}^{s/v}$ allow us to deduce the type of long-range order in either  
the spin or the valley degrees of freedom, cf. Figs.~\ref{Fig:PhaseDiagramTriangular} and~\ref{Fig:PhaseDiagramHoneycomb}.
 
\paragraph{Emergent spin-valley liquid behavior.} 

We begin our analysis with the SU(4) symmetric point, $J_{s} / J = J_{v} / J = 1$. 
For both the triangular and honeycomb lattice, no instabilities are detected in the pf-FRG flow, indicating a fully symmetric ground state. In addition, upon varying the vertex range $L$ we observe no finite-size dependence of the flows, consistent with a ground state without symmetry-breaking long-range order (see~\cite{SM} for details on the finite-size analysis). This rules out not just magnetically ordered states, but also valence bond or dimer crystals \footnote{Generally, an instability can be observed for any (quantum) order which is captured by a fermionic bilinear operator, see Ref.~\cite{PhysRevB.97.064415}. More complicated order whose order parameter involves higher order operators does not necessarily manifest in a flow breakdown}, an ordering which spins in the self-conjugate representation are often prone to \cite{AffleckMarston1989,Rokhsar1990}. For SU(4) spins in the self-conjugate representation we can further use the Lieb-Schultz-Mattis-Hastings~\cite{LIEB1961407,PhysRevB.69.104431,Yao:2018kel} theorem to rule out a featureless Mott insulator as the ground state in the case of the triangular lattice, whereas such a state is in principle still a possibility on the honeycomb lattice. We note that the spin/valley structure factors have features resembling $120^{\circ}$/N\'eel order, albeit significantly broadened, see Figs.~\ref{Fig:PhaseDiagramTriangular}(c) and \ref{Fig:PhaseDiagramHoneycomb}(c). 

\begin{figure}[b]
	\centering
	\includegraphics[width = 1.0\columnwidth]{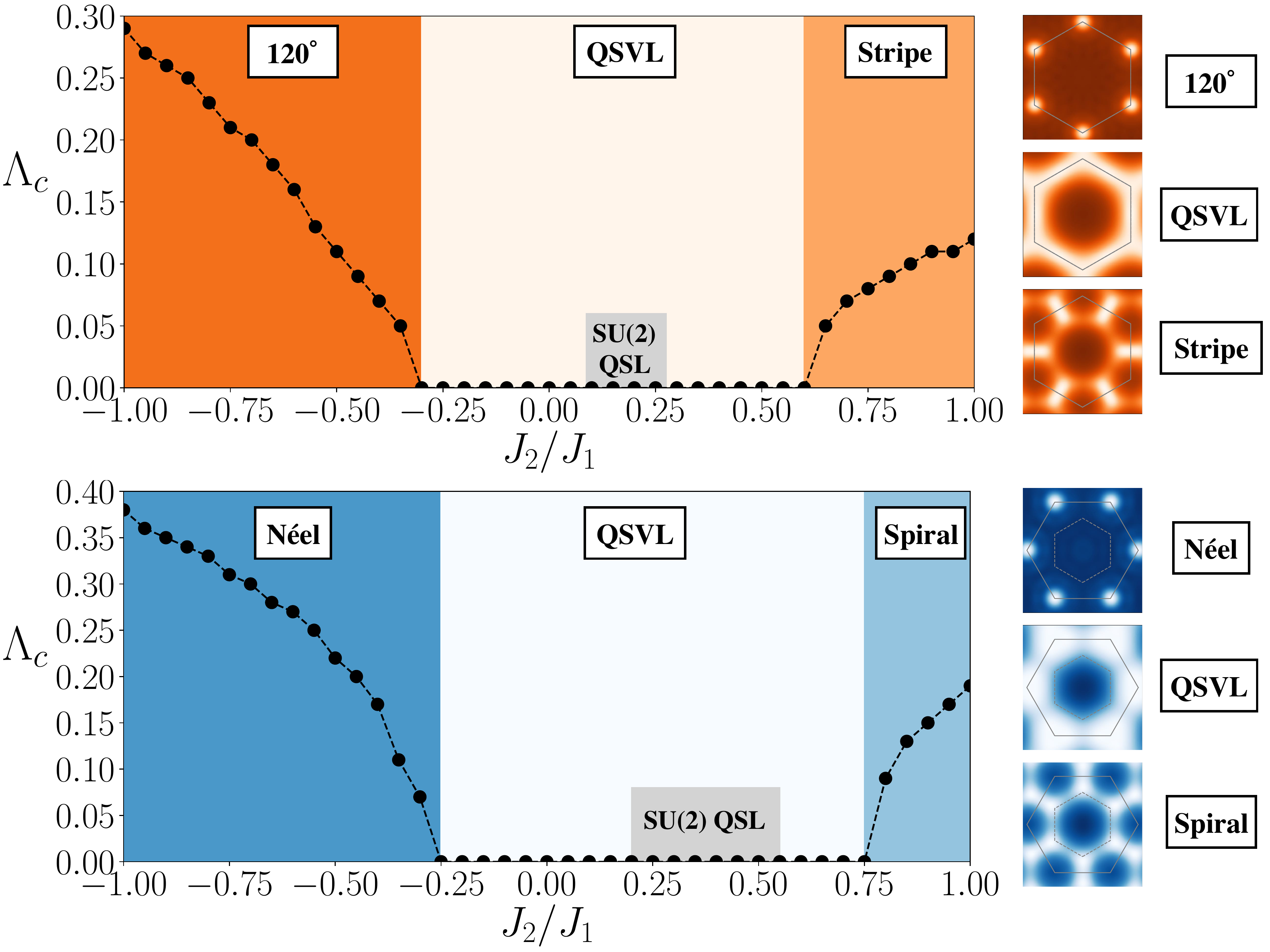}
	\caption{{\bf Phase diagrams for SU(4) $J_{1}$-$J_{2}$ models} on the triangular lattice (top) and the honeycomb lattice (bottom). Grey boxes indicate the extent of the quantum spin liquid (QSL) regime for the respective spin-1/2 SU(2) model. Structure factors for the respective phases are shown to the right, where the same color scale is applied to all plots of the underlying lattice.}
	\label{J2_01}
\end{figure}

\paragraph{Stability of spin-valley liquid and adjacent magnetism.} 

Moving towards parameter regimes with broken SU(4) symmetry, $J_{s} / J , J_{v} / J\neq 1$, 
we find that an extended paramagnetic region emanates from the SU(4) symmetric point, 
see the white wedges in Figs.~\ref{Fig:PhaseDiagramTriangular} and~\ref{Fig:PhaseDiagramHoneycomb}.
Importantly, this finding supports the stability of the emergent spin-valley liquid behavior even in the presence of SU(4) breaking perturbations such as the Hund's coupling. 
Comparing the two lattice geometries, the triangular lattice gives rise to a parametrically larger QSVL phase
than the bipartite honeycomb lattice, which can likely be traced back to the geometric frustration of the former. 
Along the diagonal line of equal coupling $J_{v}=J_{s}$, the QSVL region eventually collapses and disappears, being replaced by long-range 
antiferromagnetic order. Moving along the dotted diagonal line in the respective phase diagrams we observe a strongly suppressed breakdown scale $\Lambda_c$, relative to the surrounding parameter space, indicating that quantum fluctuations are strongest when $J_{v}=J_{s}$.

For sufficiently strong dominance of either spin or valley coupling, different ordered phases occur for both lattice geometries. 
The transition towards an ordered state is indicated by a leading instability in the RG flow, either in the spin or valley sector.
To explore the subleading instabilities in the remaining sector, we employ a heuristic mean-field-like approach to estimate the effective spin or valley couplings between nearest-neighbor sites $i$ and~$j$,
\begin{align}\label{eq:jeff}
J_{v}^{\text{eff}} = J_{v} + J\! \cdot\! \chi_{ij}^{s \Lambda_{c}}\,,\ \mathrm{and}\ \  
J_{s}^{\text{eff}} = J_{s} + J\! \cdot\! \chi_{ij}^{v \Lambda_{c}}\,.
\end{align}
Note that for $120^{\circ}$ or N\'eel order in one of the SU(2) sectors the corresponding nearest-neighbor correlation becomes negative. Therefore, the effective couplings $J_{v}^{\text{eff}}$ and $J_{s}^{\text{eff}} $ may, too, turn negative and drive
a {\em ferromagnetic} instability in the other sector, despite the antiferromagnetic nature of all couplings in the microscopic spin-valley model
\cite{wu2019ferromagnetism}.
This kind of mechanism may be at the origin of the spin polarization observed at half-filling in TDBG~\cite{liu2019spin,cao2019electric}, as first pointed out in Ref.~\cite{wu2019ferromagnetism} for quarter-filling.
Extracting the sign of the effective coupling according to Eq.~\eqref{eq:jeff} at the transition scale of the leading sector allows us to distinguish two regimes with either ferro- or antiferromagnetic correlations in the subleading sector \footnote{
We finally note that, in principle, the FRG setup can also be extended to allow for a continuation of the flow into the ordered regime~\cite{10.1143/PTP.112.943,PhysRevB.97.064416,roscher2019cluster}, however, such modifications are beyond the scope of this work. 
}. In Figs.~\ref{Fig:PhaseDiagramTriangular} and~\ref{Fig:PhaseDiagramHoneycomb} the so-determined order in the subleading regimes is indicated by  triangle  (ferromagnetic) or square  (antiferromagnetic) symbols.
%

\begin{figure}[t]
	\centering
	\includegraphics[width =\columnwidth]{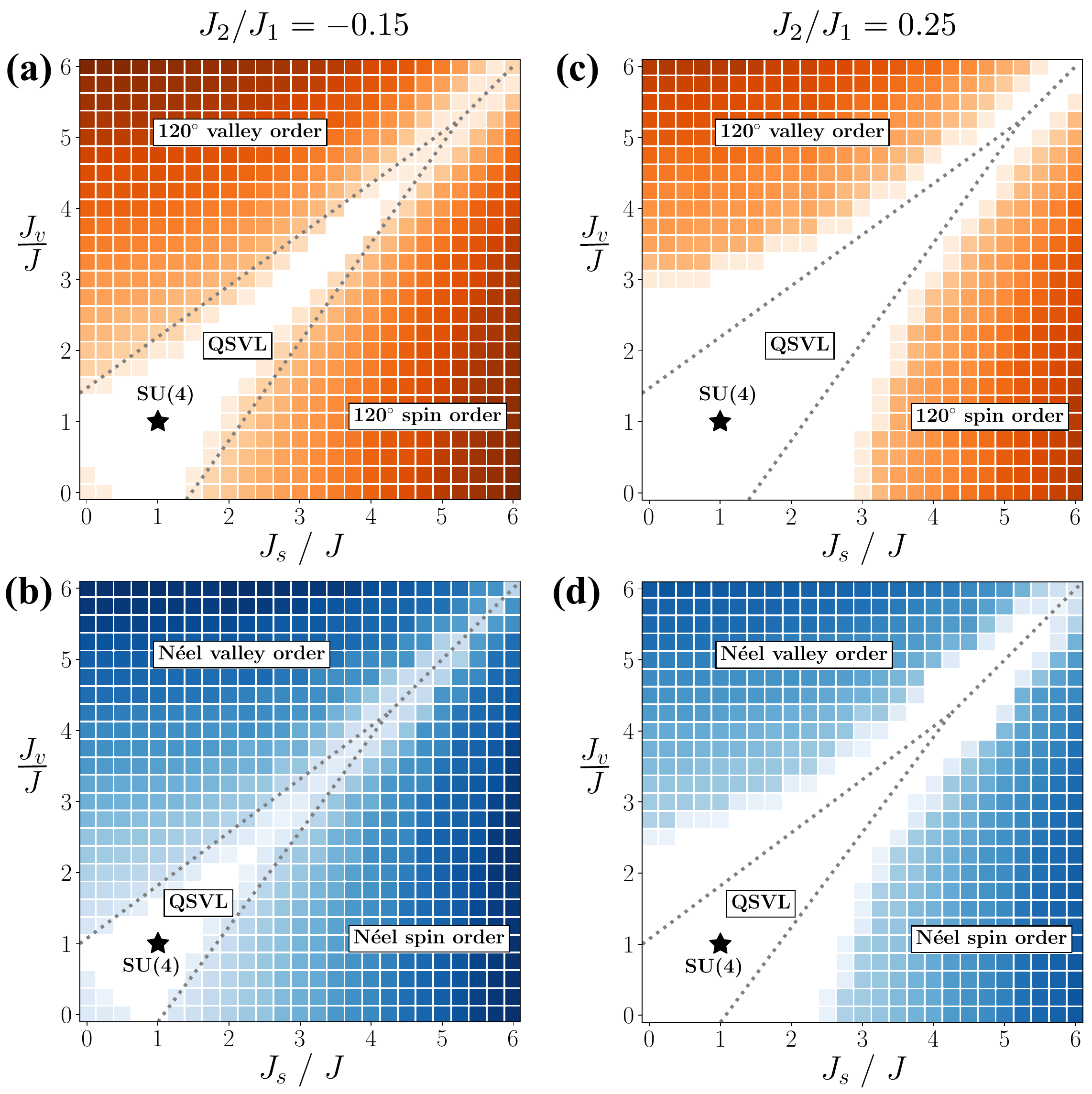}
        	\caption{{\bf Phase diagrams for longer-ranged spin-valley model} showing the effect of ferromagnetic $J_2/J_1 = -0.15$ for (a) the triangular and (b) the honeycomb model. The same for antiferromagnetic $J_2/J_1 = 0.25$ for (c) the triangular  and (d) the honeycomb model.
	Colors correspond to critical scales as indicated in Figs. \ref{Fig:PhaseDiagramTriangular} and \ref{Fig:PhaseDiagramHoneycomb}. 
	For reference, the phase boundaries at $J_2/J_1=0$ and the diagonal are marked by dotted lines.
	}
	\label{J2_02}
\end{figure}

\paragraph{Longer-range interactions.}

In the ongoing search for an effective microscopic description for moir\'e heterostructures it has been pointed out that longer-ranged Coulomb interactions should not be neglected \cite{PhysRevX.8.031087}, which in the effective spin model will give rise to exchange couplings beyond nearest-neighbor. To probe the stability of the QSVL regime in our model
we here consider the effect of a next-nearest neighbor coupling $J_2$.

Let us first recapitulate the effect a next-nearest-neighbor coupling $J_2$ for the spin-1/2 SU(2) case on the triangular and the honeycomb lattices.
Here the bare nearest neighbor coupling leads to magnetic ordering and only an antiferromagnetic $J_2$ of intermediate coupling strength facilitates the formation of a narrow quantum spin liquid (QSL) regime~\cite{PhysRevB.92.041105,PhysRevB.92.140403}, as indicated by the grey boxes in Fig.~\ref{J2_01}. Notably, the induced
QSL regime is somewhat larger for the honeycomb lattice where the next-nearest neighbor interaction introduces geometric frustration.

For the model at hand, we first concentrate on the SU(4) symmetric point and explore the effect of $J_2/J_1\in [-1,1]$. As shown in Fig.~\ref{J2_01}, the QSVL region for the SU(4) model is significantly expanded for both lattice geometries in comparison to the SU(2) QSL case.
The impact of $J_2$ on the the full spin-valley $(J_s, J_v)$ phase diagrams of Figs.~\ref{Fig:PhaseDiagramTriangular} and \ref{Fig:PhaseDiagramHoneycomb} is illustrated in Fig.~\ref{J2_02} for both ferromagnetic and antiferromagnetic $J_2$. While an antiferromagnetic  $J_2$ is found to further widen the wedge-shaped QSVL region, the converse occurs for ferromagnetic  $J_2$, which drives the system closer to the  ordered states. This means that, depending on the sign of $J_2$, longer-range interactions can actually stabilize and even expand the region of QSVL behavior.

\paragraph{Conclusions.} 

In this work, we studied $\mathrm{SU(2)}^{\rm s}\otimes \mathrm{SU(2)}^{\rm v}$-symmetric spin-valley Heisenberg models in the self-conjugate representation for both the triangular and honeycomb lattice. Seen as the effective 
Hamiltonians generated in the strong-coupling limit of an underlying Hubbard model, such models are relevant as minimal models in the exploration of the correlated insulating states of recently synthesized moir\'e heterostructures, including TBG (honeycomb) or TDBG and TLG/h-BN (triangular).
Depending on which set of minibands the Hubbard model is designed to describe, the half-filling case studied here can potentially describe different candidate correlated insulators~\cite{SM}, e.g.~the insulator at half-filling $n=+n_s/2$ in the triangular system TDBG or the honeycomb system TBG at charge neutrality $n=0$. 

In particular, we focused on the study of Hund's-induced as well as longer-ranged exchange couplings and their impact on the spin-valley liquid  which has been found to emerge in the limit of SU(4) symmetry in both lattice 
geometries.
We find extended parameter regimes where this phase is stabilized, with no signatures of long-range  order, providing evidence for a stable realm of spin-valley liquid behavior. Experimentally, such a phase would be consistent with a correlated insulator lacking spin and valley polarisation. However, the precise nature of the phase and potential experimental fingerprints are left for future study, though we note that a recent projective-symmetry-group classification of fermionic partons 
on the half-filled triangular lattice suggests the possibility of a U(1) spin liquid with four Fermi 
surfaces \cite{2019arXiv190610132Z}, which would be consistent with our analysis. 
Our findings hint at the possibility of spin-valley entangled quantum liquids lurking within the correlated insulating regimes of moir\'e heterostructures. 

\paragraph{Acknowledgments.} 

We thank L.~Classen, P. Corboz, A. Laeuchli, and A. Paramekanti for discussions. C.H. thanks the Aspen Center for Theoretical Physics, where parts of this paper were written, for hospitality and support under Grant No. NSF 1066293. 
We acknowledge partial support from the Deutsche Forschungsgemeinschaft (DFG, German Research Foundation), Projektnummer 277146847 -- SFB 1238 (project C03) and Projektnummer 277101999 -- TRR 183 (projects B01 and A02). The numerical simulations were performed on the CHEOPS cluster at RRZK Cologne and the JURECA Booster at the Forschungszentrum Juelich.

\bibliography{Bib_moire}

\clearpage
\appendix

\onecolumngrid
\section{I. Hexagonal moir\'e structures} 

As noted in the main text, the minimal model that covers the necessary universal aspects of the various moir\'e heterostructures is a two-orbital extended Hubbard model. With four flavors of fermions per site, two spin and two valley degrees of freedom, this leads to a four band model on the triangular lattice and an eight band model on the honeycomb lattice (where the doubling is simply due to the doubling of the unit cell). Which of these lattices is appropriate to use depends on the particular 
moir\'e heterostructure one is interested in.

\begin{figure}[h]
  \centering
  \includegraphics[width = 0.65\columnwidth]{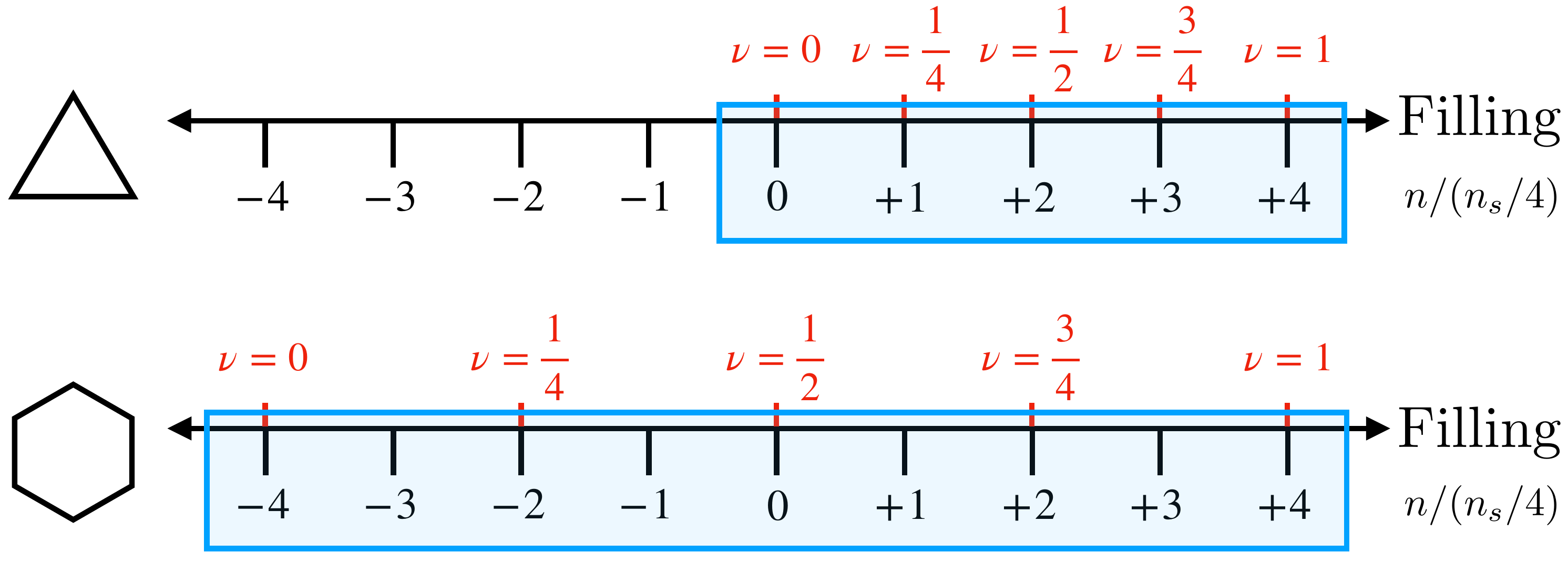}
  \caption{{\bf Comparison of filling} for the effective Hubbard model, $\nu$, and for the minibands in the experimental cases of interest, $n/(n_s/4)$ (where $n=\pm n_s$ corresponds to fully empty or filled bands) in the case of a triangular (upper) and honeycomb (lower) lattice description. Depending on the lattice description the effective Hubbard model can either apply to all of, or just half of, the minibands, with its region of applicability denoted by the blue boxes.}
  \label{Fig:Filling}
\end{figure}

For TBG, TLG/h-BN and TDBG there are a total of eight minibands near charge neutrality, four above and four below, that are separated from the rest of the spectrum by trivial band gaps. Filling of these minibands is thus typically denoted as ranging from $n=-n_s$ to $n=+n_s$, as indicated in Fig.~\ref{Fig:Filling} (where, for convenience, we plot $n/(n_s/4)$). In the case of TBG, the bands above/below charge neutrality are connected via Dirac points, meaning that any effective Hubbard model must describe all eight bands. This naturally motivates the use of the honeycomb lattice Hubbard model. Half-filling, i.e. the scenario focused on in the main text, thus corresponds to charge neutrality $n=0$. On the other hand, in the case of TDBG and TLG/h-BN the bands above/below charge neutrality are disconnected from one another, meaning that an effective Hubbard model description need only focus on one or the other set of four bands. This naturally leads to a triangular lattice description, with half-filling now corresponding to $n=\pm n_s/ 2$. 

\section{II. Pseudofermion functional RG approach}

The pseudofermion functional renormalization group (pf-FRG) has recently been established as a versatile tool for the investigation of ground state phase diagrams for a wide class of spin models~\cite{PhysRevB.81.144410,WETTERICH199390}. 
In doing so, the free fermion propagator $G_0=(i\omega)^{-1}$ of a pseudofermion decomposed quartic Hamiltonian, e.g., Eq.~\eqref{eq:H}, is modified by a step-like regularization function $\Theta(|\omega|-\Lambda)$ with frequency cutoff scale~$\Lambda$, i.e. $G_0 \to G_0^\Lambda=G_0\Theta^{\Lambda}$. 
The artificial scale dependence of this theory results in a hierarchy of
coupled one-loop RG flow equations for the one-particle-irreducible~(1PI) interaction vertices. We employ a standard approximation scheme, where the hierarchy is truncated to exclusively account for the frequency-dependent self-energy $\Sigma^\Lambda$ and two-particle interaction vertex $\Gamma^\Lambda$, see, e.g., Ref.~\cite{PhysRevB.97.064415} for more details and technicalities.

Here, we describe the aspects of the pf-FRG which are particular to the present spin-valley model, i.e. the vertex parametrization for the $\mathrm{SU(2)}\otimes\mathrm{SU(2)}$ symmetry and the implementation of the filling constraint.

\subsubsection{II.~A. Vertex parametrization for SU(2)$\otimes$SU(2) symmetry}

The pseudofermion decomposition of the spin-valley operators $\hat \sigma^a_i,\hat \tau^b_i,\hat \sigma^a_i\otimes \hat \tau^b_i$ in the Hamiltonian, Eq.~\eqref{eq:H}, exhibits a local U(1)~symmetry.
Consequentially, the self-energy $\Sigma^\Lambda$ can be efficiently parametrized as being local and the two-particle interaction vertex $\Gamma^\Lambda$ as being bilocal. 
Translation invariance in imaginary time additionally reduces the number of independent frequency arguments by one. 
The spin/valley dependence of the 1PI irreducible vertices can be expanded in terms of an $\mathfrak{su}(2)$ basis. 
This scheme is augmented by symmetry-allowed SU(2)-invariant density terms with the most general parametrization reading
\begin{align}
\Sigma^{\Lambda}(1';1) &\!=\!\sum_{\alpha, \beta} \Sigma^{\Lambda \alpha \beta}_{i_{1}}(w_{1}) \theta^{\alpha}_{s_{1'} s_{1}} \theta^{\beta}_{l_{1'} l_{1}} \delta_{i_{1'} i_{1}} \delta(w_{1'} - w_{1}) \,, \\ \notag
\Gamma^{\Lambda}(1',2';1,2) &\!=\!
\sum_{\substack{\alpha, \alpha' \\ \beta, \beta'}} \Gamma^{\Lambda \alpha \alpha' \beta \beta'}_{i_{1}i_{2}}(w_{1'} w_{2'}; w_{1} w_{2}) \theta^{\alpha}_{s_{1'} s_{1}} \theta^{\alpha'}_{s_{2'} s_{2}} \theta^{\beta}_{l_{1'} l_{1}} \theta^{\beta'}_{l_{2'} l_{2}} \delta_{i_{1'} i_{1}} \delta_{i_{2'} i_{2}} \delta(w_{1'} + w_{2'} - w_{1} - w_{2}) - (1 \leftrightarrow 2) \,,
\end{align}
where $1 = \{ i_{1}, s_{1}, l_{1}, w_{1} \}$ and $\alpha, \beta \in \{0, 1, 2, 3\}$ with $\theta^{0} = \mathbf{1}$. Exploiting SU(2)~symmetry in both spin and valley indices we are left with pure density contributions for the self-energy, while the two-particle vertex may also contain off-diagonal terms albeit with equal spin directions, i.e.
\begin{align}
\Sigma^{\Lambda}(1'; 1) &= \Sigma^{\Lambda}_{i_{1}}(w_{1}) \delta_{s_{1'} s_{1}} \delta_{l_{1'} l_{1}} \delta_{i_{1'} i_{1}} \delta(w_{1'} - w_{1}) \,, \\ \notag
\Gamma^{\Lambda}(1', 2'; 1, 2) &= \bigg{[} \Gamma^{\Lambda ss}_{i_{1}i_{2}}(w_{1'} w_{2'}; w_{1} w_{2}) \theta^{a}_{s_{1'} s_{1}} \theta^{a}_{s_{2'} s_{2}} \theta^{b}_{l_{1'} l_{1}} \theta^{b}_{l_{2'} l_{2}} + \Gamma^{\Lambda sd}_{i_{1}i_{2}}(w_{1'} w_{2'}; w_{1} w_{2}) \theta^{a}_{s_{1'} s_{1}} \theta^{a}_{s_{2'} s_{2}} \delta_{l_{1'} l_{1}} \delta_{l_{2'} l_{2}} \\ \notag
&\phantom{\bigg{[}} +\Gamma^{\Lambda ds}_{i_{1}i_{2}}(w_{1'} w_{2'}; w_{1} w_{2}) \delta_{s_{1'} s_{1}} \delta_{s_{2'} s_{2}} \theta^{b}_{l_{1'} l_{1}} \theta^{b}_{l_{2'} l_{2}} +\Gamma^{\Lambda dd}_{i_{1}i_{2}}(w_{1'} w_{2'}; w_{1} w_{2}) \delta_{s_{1'} s_{1}} \delta_{s_{2'} s_{2}} \delta_{l_{1'} l_{1}} \delta_{l_{2'} l_{2}} \bigg{]} \\ 
&\phantom{\bigg{[}} \times \delta_{i_{1'} i_{1}} \delta_{i_{2'} i_{2}} \delta(w_{1'} + w_{2'} - w_{1} - w_{2}) - (1 \leftrightarrow 2) \,,
\end{align}
where $a, b \in \{1, 2, 3\}$. 

The initial conditions at the UV scale then read $\Sigma^{\infty}_{i_{1}}(w) = 0$ for the self-energy and 
\begin{align}
&\Gamma^{\infty ss}_{i_{1}i_{2}}(w_{1'} w_{2'}; w_{1} w_{2}) = J_{\phantom{v}}, \
 \Gamma^{\infty sd}_{i_{1}i_{2}}(w_{1'} w_{2'}; w_{1} w_{2}) = J_{s}\,, \\ \notag 
&\Gamma^{\infty ds}_{i_{1}i_{2}}(w_{1'} w_{2'}; w_{1} w_{2}) = J_{v}, \
 \Gamma^{\infty dd}_{i_{1}i_{2}}(w_{1'} w_{2'}; w_{1} w_{2}) = 0\,,
\end{align}
for the two-particle interaction vertices.

\subsubsection{II.~B. Particle-hole symmetry and the half-filling constraint}

The pf-FRG approach considered here explicitly implements the particle-hole symmetry of the spin-valley model on the level of the 1PI vertices. 
As a consequence, the half-filling constraint is naturally fulfilled on average upon setting the chemical potential to zero and local fluctuations around the average are energetically penalized~\cite{PhysRevB.81.144410}.
Therefore, contributions from unphysical sectors of the Hilbert space violating the local half-filling constraint are strongly suppressed at zero temperature.
Consequently, the local half-filling constraint is expected to be well-enforced within the pf-FRG approach as long as the flow does not develop an instability.

Specifically, in the model studied here, the local Hilbert space for fermionic flavors $\alpha \in  \{(\uparrow\!+),(\uparrow\!-),(\downarrow\!+),(\downarrow\!-)\}$ is equipped with the particle-number basis $B = \{|n_{1}, ..., n_{4} \rangle \}$. We define the linear unitary operator $\mathcal{P}$ acting on the basis by exchanging each occupied state with an empty state $\mathcal{P} |n_{1}, ..., n_{4} \rangle = | 1 - n_{1}, ..., 1 - n_{4} \rangle$.
By computing the corresponding matrix elements, one finds that $\mathcal{P}$ transforms creation and annihilation operators into each other, i.e
$\mathcal{P}^{\dagger} f^{\dagger}_{\alpha} \mathcal{P}^{\phantom{\dagger}}\! =\!f^{\phantom{\dagger}}_{\alpha}\,,\
\mathcal{P}^{\dagger} f^{\phantom{\dagger}}_{\alpha} \mathcal{P}^{\phantom{\dagger}}\!=\! f^{\dagger}_{\alpha}$. This transformation leaves the spin-valley Hamiltonian invariant. On the level of vertex functions, we obtain the identities $\Sigma^{\Lambda}(1'; 1) = -\Sigma^{\Lambda}(-1; -1')$ and $\Gamma^{\Lambda}(1', 2'; 1, 2) = \Gamma^{\Lambda}(-1, -2; -1', -2')$, where the minus sign applies only to Matsubara frequencies. The vertex components therefore obey $\Sigma^{\Lambda}_{i_{1}}(w)= -\Sigma^{\Lambda}_{i_{1}}(-w)$ and $\Gamma^{\Lambda \zeta}_{i_{1} i_{2}}(s, t, u)= \Gamma^{\Lambda \zeta}_{i_{1} i _{2}}(-s, -t, -u)$ where $\zeta \in \{ss, sd, ds, dd\}$.

These symmetries are explicitly implemented in our pf-FRG approach. However, only one local subspace, namely the one with two occupied states, is mapped to itself by $\mathcal{P}$, i.e. by enforcing the symmetries of that respective subspace we can be sure that the RG flow  starts from half-filling at each lattice site. Furthermore, since the particle number per site must be conserved as a consequence of local U(1) symmetry, hopping processes that alter the filling would trigger a measurable non-magnetic instability of the flow, which we do not observe here.

\subsection{II.~C. Finite-size analysis of the RG flow}

\begin{figure}[b!]
  \centering
  \includegraphics[width = 0.48\columnwidth]{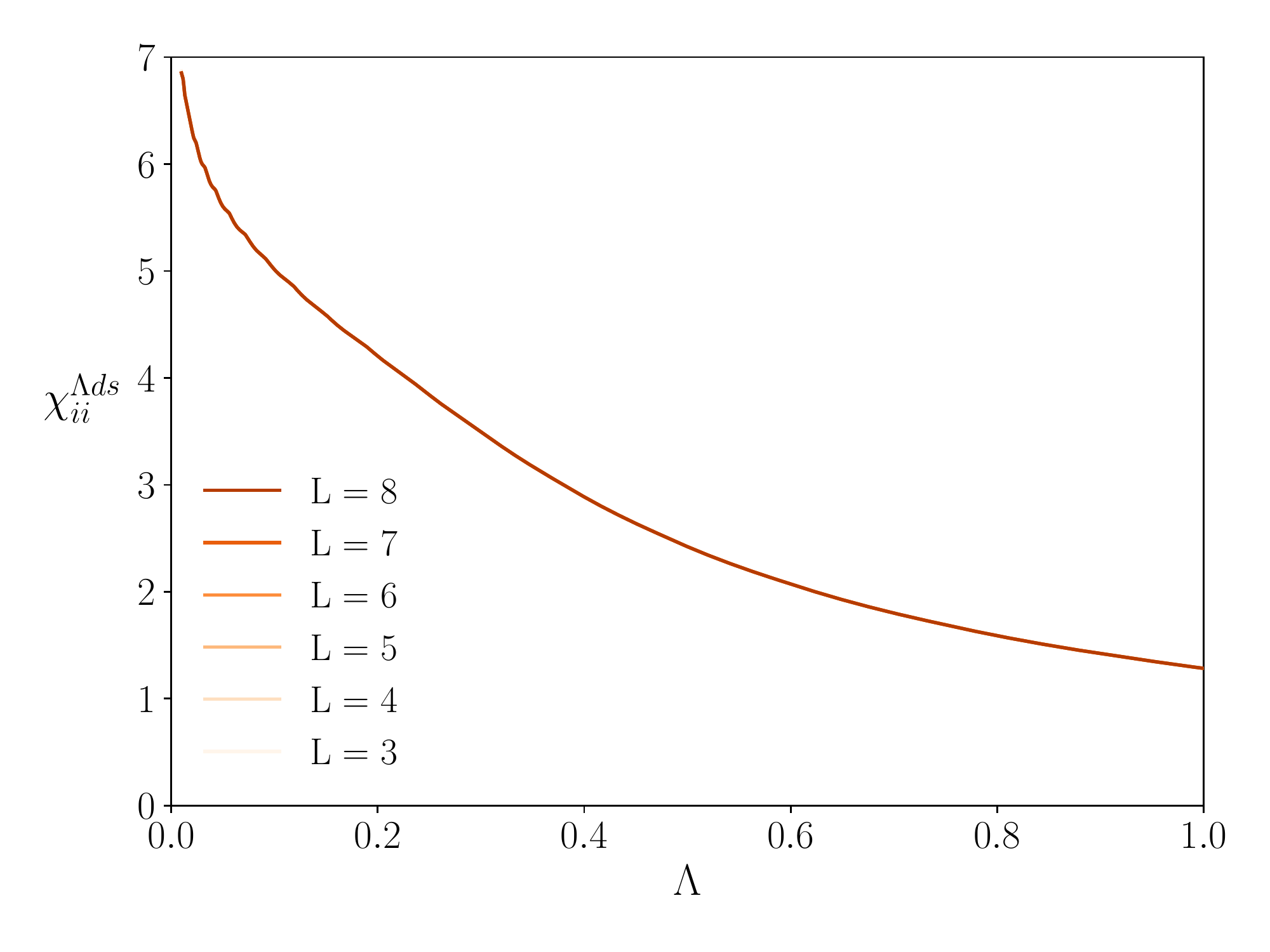}
  \includegraphics[width = 0.48\columnwidth]{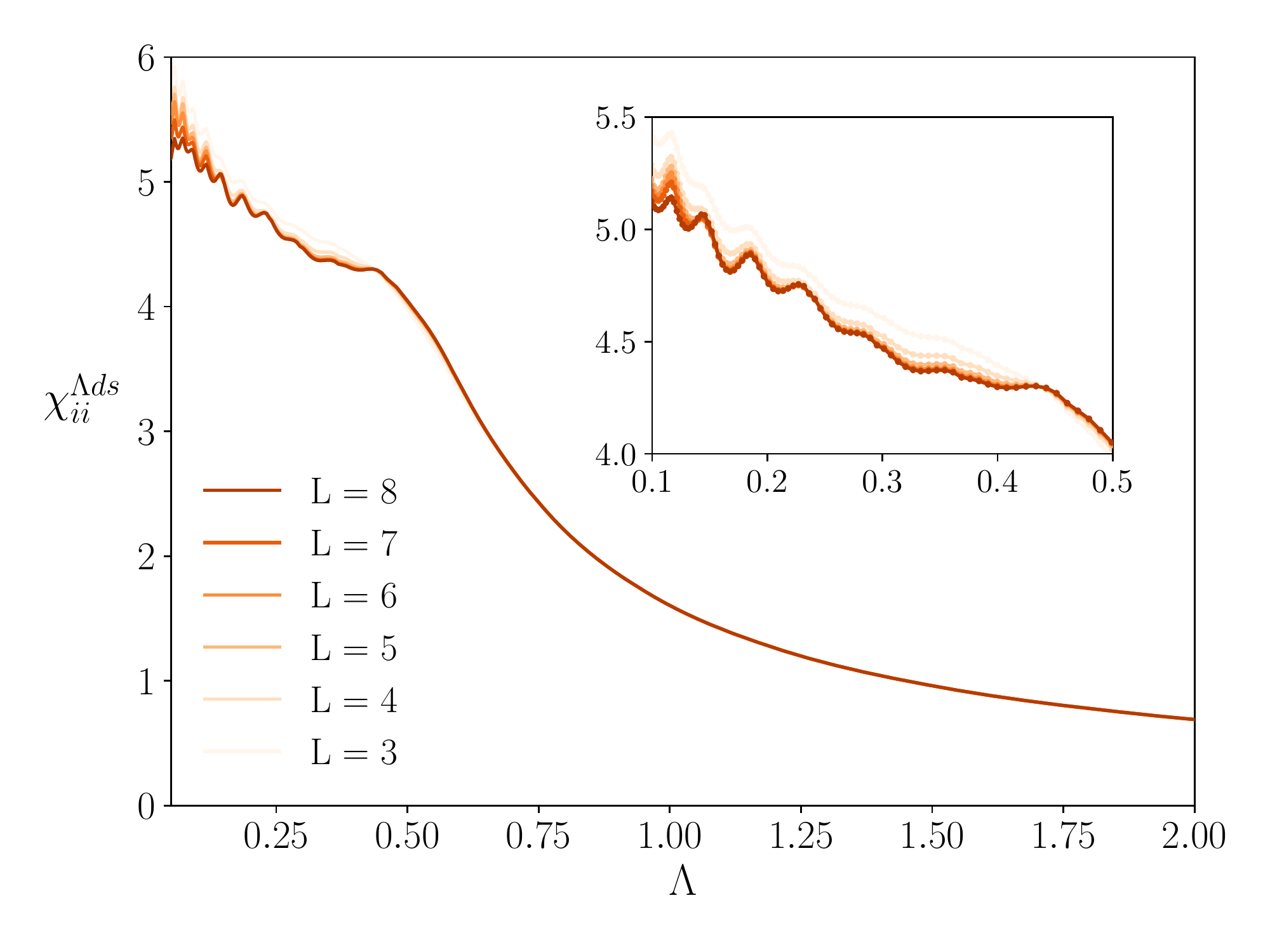}
  \caption{{\bf Finite-size analysis} of the onsite correlation function $\chi^{\Lambda ds}_{ii} = \chi_{ii}^{v \Lambda}$ for the {\bf triangular lattice}. 
  		Left: $J_{s} / J = J_{v} / J = 1.0$; Right: $J_{s} / J = 0.5, J_{v} / J = 4.0$. 
		For a paramagnetic ground state the flow shows neither dependence on $L$ nor an instability.}
  \label{Scaling Triangular}
\end{figure}

\begin{figure}[b!]
  \centering
  \includegraphics[width = 0.48\columnwidth]{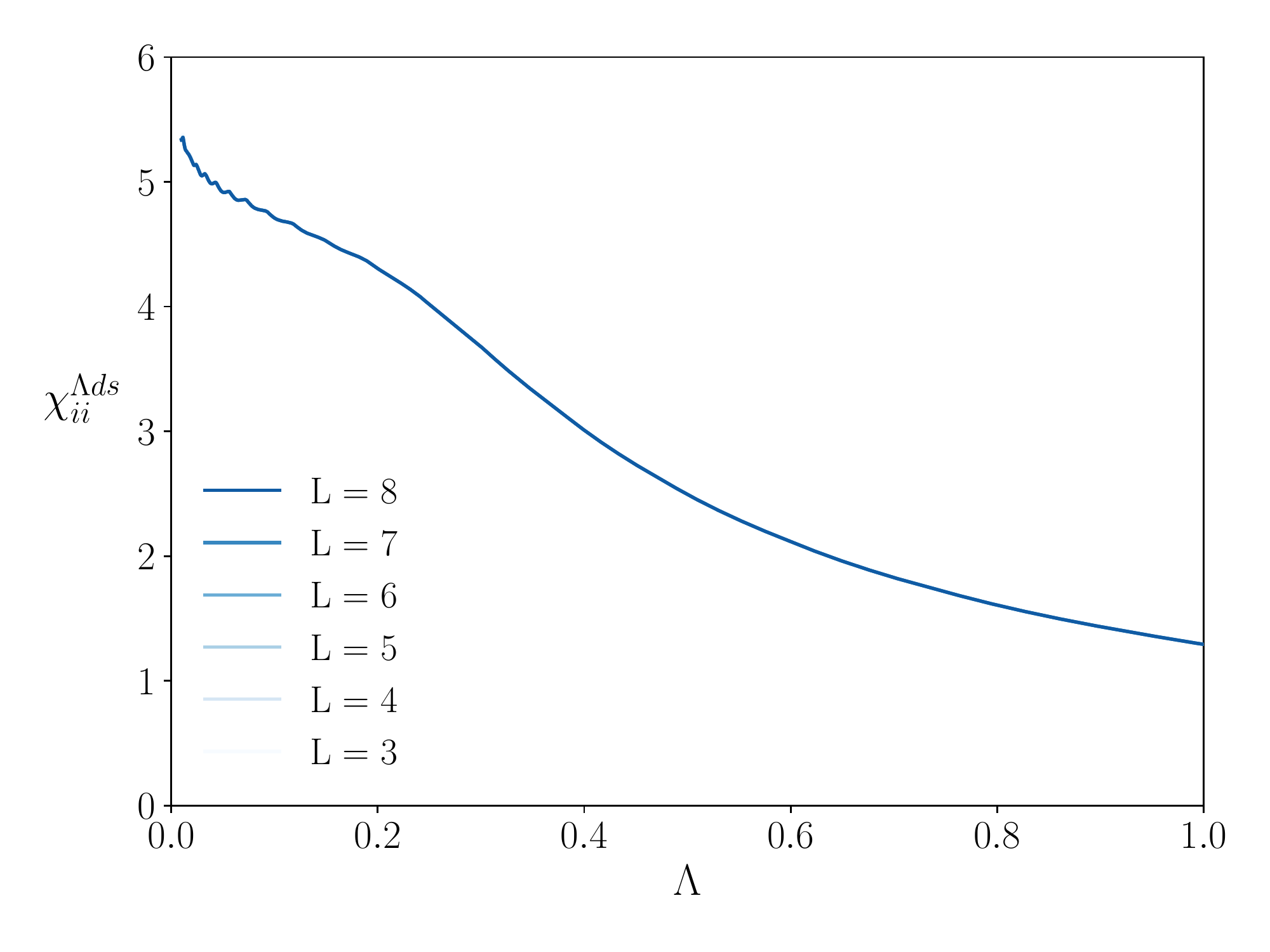}
  \includegraphics[width = 0.48\columnwidth]{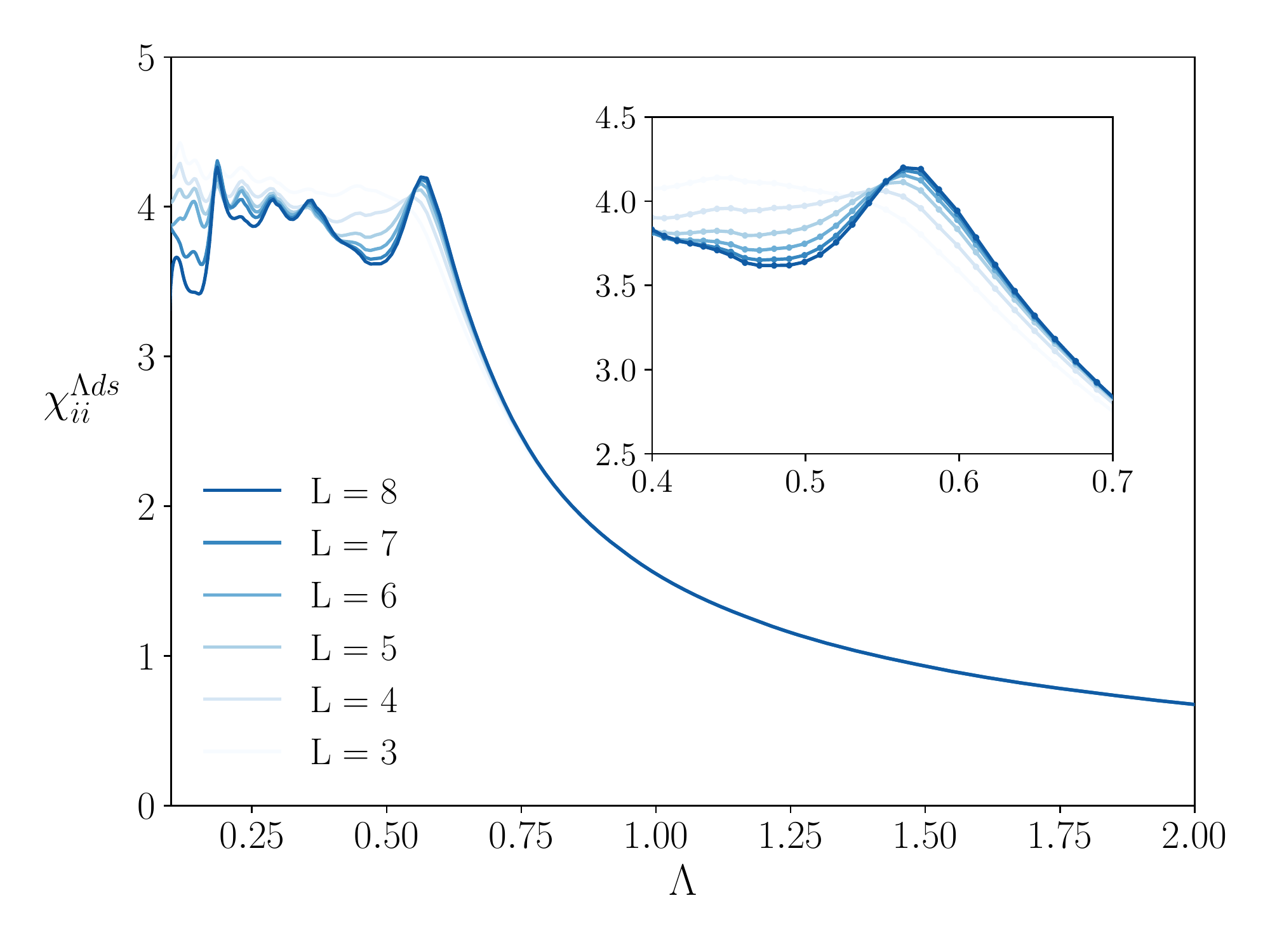}
  \caption{{\bf Finite-size analysis} of the onsite correlation function  $\chi^{\Lambda ds}_{ii} = \chi_{ii}^{v \Lambda}$ for the {\bf honeycomb lattice}. 
  Left: $J_{s} / J = J_{v} / J = 1.0$; Right: $J_{s} / J = 0.5, J_{v} / J = 4.0$. 
  For a paramagnetic ground state the flow shows neither dependence on $L$ nor an instability.}
  \label{Scaling Honeycomb}
\end{figure}

An instability in the vertex function during the RG flow indicates a spontaneous breaking of symmetries that have been implemented in the initial conditions~\cite{PhysRevB.97.064415}. 
Most prominently, magnetic instabilities appear as pronounced kinks or cusps in the flow of the momentum resolved two-spin correlations.
Alternatively, one may check the behavior of an on-site correlation function, i.e., $\chi^{\Lambda}_{ii}$, for different values of the vertex range $L$.
Formally, $L$ does not determine the system size (which is in fact infinite in pf-FRG) but rather sets the scale on which spins can be correlated. It is then natural to expect sensitivity to this parameter near the critical scale since the physics is governed by the collective behavior of all spins. On the other hand, if the system does not  develop an instability down to the smallest energy scales, i.e. the pf-FRG flow stays regular, real space correlations should be robust with respect to variations of $L$. 

Indeed as shown in Figs. \ref{Scaling Triangular} and \ref{Scaling Honeycomb}, flows of the spin correlation in the dominant interaction channel for different $L$ are aligned within the paramagnetic regions of the spin-valley phase diagrams, but deviate from each other around the critical scale in the ordered phases. We find, however, that this effect is more subtle for the triangular than the honeycomb  lattice, which we attribute to the inherent geometric frustration of the former.

\subsection{III. Structure factor evolution in the spin-valley liquid of the $J_1$-$J_2$ model}

The spin-valley entangled liquid ground states of the nearest-neighbor SU(4) Heisenberg models (on both the triangular and honeycomb lattice) remain stable upon inclusion of moderate longer-ranged exchange interactions as illustrated in Fig.~\ref{J2_01} of the main text.

\begin{figure}[h!]
	\centering
	\includegraphics[width=\columnwidth]{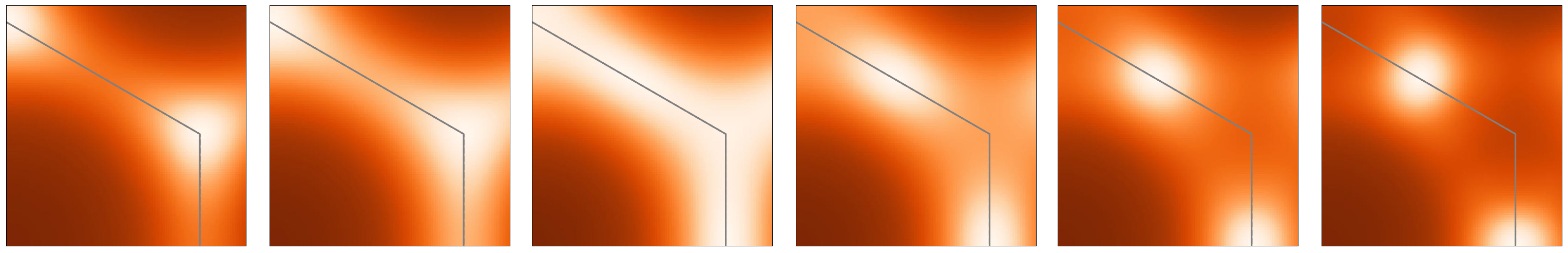}
	\caption{{\bf  Structure factors  for the SU(4) model on the triangular lattice} within the paramagnetic phase. Lines denote the first Brillouin zone. From left to right $J_{2} / J_{1} = 0.0/ 0.1/ 0.2/ 0.3/ 0.4/ 0.5$. At a ratio of $J_{2} / J_{1} \approx 0.2$ a deformation from local $120^{\circ}$ to local stripe correlations is observed.}
	\label{SU4_Triangular}
\end{figure}

\begin{figure}[h!]
	\centering
	\includegraphics[width=\columnwidth]{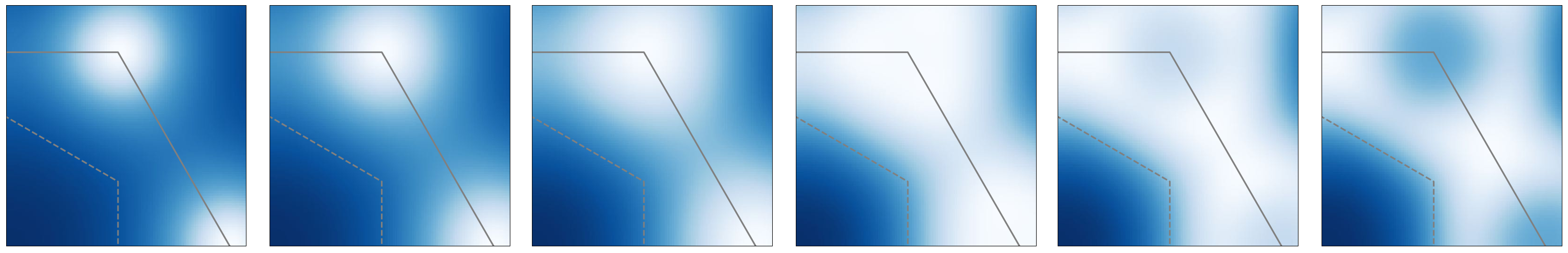}
	\caption{{\bf Structure factors for the SU(4) model on the honeycomb lattice} within the paramagnetic phase. Dashed lines denote the first, full lines the extended Brillouin zone. From left to right $J_{2} / J_{1} = 0.0/ 0.1/ 0.2/ 0.3/ 0.4/ 0.5$. At a ratio of $J_{2} / J_{1} \approx 0.3$ a deformation from local N\'eel to local spiral correlations is observed.}
	\label{SU4_Honeycomb}
\end{figure}

Here, we provide further information about the evolution of the structure factors upon varying $J_2/J_1$.
First, we recall that for $J_2 / J_1 = 0$, local correlations are reminiscent of $120^{\circ}$ (N\'eel) order for the triangular (honeycomb) model.
Going to large antiferromagnetic $J_2>0$, stripe (spiral) order emerges with the evolution of the structure factor being plotted in Figs. \ref{SU4_Triangular} and \ref{SU4_Honeycomb} at the onset of these orders.
Around $J_2 / J_1 \approx 0.2$ for the triangular and at about $J_2 / J_1 \approx 0.3$ for the honeycomb lattice, the topology of the momentum resolved correlation functions changes visibly, indicating a Lifshitz transition.

\end{document}